\documentclass[]{aastex631}
\usepackage{hyperref}

\newcommand{\beq}{\begin{eqnarray}}
\newcommand{\eeq}{\end{eqnarray}}

\usepackage{amssymb}
\begin{document}

\title{Afterglows from binary neutron star post-merger systems embedded in AGN disks}

\author[0000-0002-8560-692X]{Adithan Kathirgamaraju}
\affiliation{Theoretical Division, Los Alamos National Laboratory, Los Alamos, NM 87545, USA}
\affiliation{CCS-2, Los Alamos National Laboratory, P.O. Box 1663, Los Alamos, NM 87545, USA}
\author[0000-0003-3556-6568]{Hui Li}
\affiliation{Theoretical Division, Los Alamos National Laboratory, Los Alamos, NM 87545, USA}
\author[0000-0001-8939-4461]{Benjamin R. Ryan}
\affiliation{CCS-2, Los Alamos National Laboratory, P.O. Box 1663, Los Alamos, NM 87545, USA}
\author[0000-0002-9182-2047]{Alexander Tchekhovskoy}
\affiliation{Center for Interdisciplinary Exploration \& Research in Astrophysics (CIERA), Physics \& Astronomy, Northwestern University, Evanston, IL 60202, USA}


\begin{abstract}

The observability of afterglows from binary neutron star mergers, occurring 
within AGN disks is investigated. We perform 3D GRMHD simulations 
of a post-merger system, and follow the jet launched from 
the compact object. 
We use semi-analytic techniques to study the propagation 
of the blast wave powered by the jet through an AGN disk-like 
external environment, extending to distances 
beyond the disk scale height. The synchrotron 
emission produced by the jet-driven forward shock is 
calculated to obtain the afterglow emission. The 
observability of this emission at different frequencies is assessed by 
comparing it to the quiescent AGN emission. 
In the scenarios where the afterglow could temporarily outshine 
the AGN, we find that detection will be 
more feasible at higher frequencies ($\gtrsim$10$^{14}$ Hz) 
and the electromagnetic counterpart could manifest as a fast variability 
in the AGN emission, on timescales less than a day.

\end{abstract}

\keywords{Magnetohydrodynamical simulations(1966) --- Radiative transfer(1335) --- High energy astrophysics(739) --- Transient sources(1851) --- Relativistic jets(1390)}


\section{Introduction} \label{sec:intro}

To date, there has been one conclusive detection of a binary 
neutron star (BNS) merger in both electromagnetic (EM) and 
gravitational waves (GW) \citep{GW170817}. A handful of 
detections either in purely GWs or EM signals have been 
loosely associated with BNS mergers (e.g., \citealt{GW190425}). 
These mergers are believed to have occurred in isolated environments. 
However another likely site for such mergers is the central regions of galaxies, where the strong gravitational potential of the 
galactic center and the central supermassive black hole (SMBH) attract an over dense population of stellar objects, whose deaths can leave behind compact objects (NSs or black holes). 
The class of galactic centers known as active galactic nuclei 
(AGNs), produce electromagnetic emission across a wide range of 
wavelengths, and is powered by the accretion of matter onto 
the SMBH which is supplied through a disk surrounding the 
SMBH (hereafter called AGN disk). Given the higher density of stars present near the centers of galaxies, a fraction of compact object mergers can be expected to occur in this central region, within an AGN disk. In fact, there have been a few 
GRBs detected recently, whose properties and location 
constraints lend evidence to a compact object merger 
origin within their galactic nucleus \citep{levan2023,lazzati2023}. 

It has been suggested that AGN disks could contain non-negligible number of stars (e.g., \citealt{syer1991,artymowicz1993}).
Broadly speaking, there are two scenarios that have 
received the most attention: the first is the in-situ formation 
scenario where AGN disks beyond some radius will 
become unstable driven by self-gravity, which could 
lead to copious amount of star formation (e.g., \citealt{levin2003,goodman2004,Dittmann2020}); 
the other is the capture scenario where the stars in 
the nuclear star cluster could interact with AGN disks and eventually get captured as stars lose energy and 
momentum (e.g., \citealt{syer1991,artymowicz1993,
macleod2020,fabj2020}). Various arguments have been 
put forth to suggest that stars in AGN disks could grow to be 
very massive (e.g., via accretion \cite{cantiello2021}) , their evolution tends to leave behind 
compact objects such as neutron stars and stellar 
mass black holes (e.g., \citealt{perna2021a,tagawa2020}). 
The subsequent joint evolution of the AGN disk, stars 
and compact objects is a subject of great interest and 
active on-going research. While the statistics of the NS-NS binary population in AGN 
disks are still uncertain, it has been suggested that rates of compact object (involving both NSs and BHs can be enchanced within AGN disks) (e.g., \citealt{mckernan2020}). 
Following previous studies (e.g., \citealt{perna2021b,zhu2021,lazzati2022}), we focus 
our effort  on the possible observational signatures 
if such BNS merger events were to occur in AGN disks. 

The GRB prompt emission and afterglow are the two 
primary non-thermal EM counterparts of a BNS, which 
are powered by a jet or possibly a cocoon inflated by the jet. 
In the case where most of the energy is imparted into 
the cocoon, a thermal shock-breakout emission will likely 
be the dominant EM counterpart (e.g., \citealt{gottlieb2018a}). 
Although the environment will play a negligible role in affecting 
the GW signals from the merger, the EM emission by
an embedded BNS merger can be 
significantly altered. This is because the high densities 
and temperatures of AGN disk environments can alter 
the dynamic evolution of the outflows from a post-BNS 
merger system (hereafter called a post-merger system). 
In addition, the photons that propagate from the outflows, 
through the AGN disks can be highly absorbed or 
scattered. In this work we focus on the longer-lived 
afterglows from embedded BNS mergers. 

Previous works have investigated the outcome of the 
EM counterparts from such embedded systems for 
the prompt emission (e.g., \citealt{lazzati2022,yuan2022,ray2023}), 
the shock breakout emission (e.g., \citealt{kimura2021,tagawa2023,zhu2021}), the
afterglow emission (e.g., \citealt{wang2022}), and emission due to the dynamical ejecta-disk interaction \citep{ren2022} . 
There have been a few numerically motivated studies on how the AGN disk material can affect the afterglow and its observability \citep{wang2022}. These works have focused on post-merger systems in AGN disks
 starting with an injection of a jet into an AGN-like 
ambient medium, rather than beginning at the source 
of the outflow, the accreting compact object. Obtaining a more accurate model for the structure 
of the jet and outflows requires initializing the system with 
an accreting compact object, that powers these outflows. 
This step is vital in modeling the EM emission as previous 
studies have stressed the importance the jet structure 
has on the observed EM counterpart \citep{marguttireview2021}.

In this work, we utilize 3D GRMHD simulations of a 
post-merger system, consisting of a black hole surrounded 
by a strongly magnetized accretion torus, and dynamical 
ejecta. We obtain the properties of the outflows from the 
simulation and extend it to distances beyond the disk 
scale height using a 2-D dynamic evolution for the blast 
wave. We calculate the synchrotron emission from this 
blast-wave, taking into account absorption by the AGN 
disks and implementing a crude treatment for scattering. 
We assess the detectability of the afterglow above the 
quiescent AGN emission. 

The paper is structured as follows: in \S  2, we 
describe the initial setup of the GRMHD simulations, 
and how the outflow is extended to beyond the AGN disk scale height. 
In \S 3, we outline how the afterglow emission is 
calculated and give the spectral evolution and light curves for select cases of ambient and blast wave parameters.
in \S 4 the detectability of the afterglow is discussed 
for a range of parameter space of the AGN disk density and scale height.
We conclude in \S 5.

\section{Initial setup and dynamics of outflow}

Here we discuss the properties of the simulations and 
initial setup of our BNS post-merger system. We explain 
how we utilize the outflow properties generated from 
the accreting compact object, and evolve it to 
distances beyond the AGN disk scale height.

\subsection{Simulation of the post-merger system}
\label{} 
Following a BNS merger, the cores of the NSs merge 
to form a compact object (a black hole for our purposes), 
surrounded by a torus of neutron star material. The final 
stages of the merger also unbinds a small fraction 
($\leq$ 0.001) of the NS from tidal interactions and 
shock ejection (e.g., \citealt{shibata2019, sekiguchi2016, 
hotokezaka2013}). This partially bound material forms 
the dynamical ejecta. The black hole, torus, and dynamical 
ejecta are the basic components of our initial setup. 

We carry out our simulation  using HARMPI\footnote{Available at
  https://github.com/atchekho/harmpi \label{fn:harmpi}}, an 
  enhanced version of the serial open-source 
code HARM \citep{gammie_2003,noble_2006}. 
This version includes additional features, e.g., taking into 
account neutrino and antineutrino emission, nuclear 
recombination and ability to track electron fraction (see \citealt{fernandez2018} for full details on these 
physical processes). The simulations are initialized 
 with an axisymmetric setup consisting of a spinning black hole encircled by a torus of 
magnetized material which is surrounded by the 
dynamical ejecta. While the initial setup is axisymmetric, the subsequent evolution of the system produces non-axisymmetric features (e.g., due to magnetorotational instabilities and turbulent behavior of the torus), which affects the accretion and outflow properties of this system. Therefore, a 3D simulation is required to accurately capture this evolution. We scale our BH mass ($M_{bh}$) to 3 solar masses 
(M$_{\odot}$) with a dimensionless spin of 0.8, the mass of the torus (as initiated following \citealt{fishbone1976}) is 0.033 M$_{\odot}$, and has an inner radius of 6 
$r_g$,  where $r_g = GM_{bh}/c^2$, and the radius of maximum pressure is 12 $r_g$. The initial torus setup is prescribed by the analytic expressions provided in \citealt{fishbone1976}, the simulation follows the detailed evolution of the torus, tracking the momentum and energy transport generated by the magnetorotational instability driven turbulence, and energy released due to nuclear recombination and neutrino emission.
The torus is seeded with a poloidal magnetic field,
governed by the vector potential $A_\phi\propto r^{5}\rho^2$ 
and has a maximum field strength of $\sim 4\times 10^{14}$ G, 
where $r$ is the radius in spherical coordinates and 
$\rho$ is density. The high magnetic field strength causes the disk to become magnetically arrested after the onset of accretion. The resolution of the grid is set at 
$768\times384\times128$ in the radial ($r$), meridional ($\theta$) 
and azimuthal ($\phi$) directions respectively, with the simulation covering a full $2\pi$ range in $\phi$. The simulation is performed in Modified Kerr-Schild coordinated with $h$ parameter of 0.3 \citep{mckinney2004}. The grid is spaced logarithmically in $r$ starting at $1.39 r_g$ (within the event horizon of the spinning BH), out to $10^5 r_g$. A higher concentration of cells is placed along the spin axis and equatorial plane, effectively doubling the resolution in these regions in order to better capture the relativistic jets and evolution of the torus.

The black hole-torus system is surrounded by dynamical 
ejecta of total mass $2\times 10^{-4}$ M$_{\odot}$\footnote{Numerical simulations indicate dynamical ejecta masses varying between $\sim 10^{-4} - 10^{-3}M_{\odot}$ \citep{radice2018}} starting 
at an inner radius of $150$ r$_{\rm g}$ with a density 
distribution that depends on radius and polar angle as
\beq
        \rho = \Lambda\left(a+\frac{b}{1+e^{-20(\theta-\frac{\pi}{4})}}\right)r^{-6},
        \label{ejectadensity}
\eeq
where $\Lambda$ is a normalization constant, obtained by fixing the total mass of the ejecta. 
From equation \ref{ejectadensity}, the ratio of densities at the equator ($\theta=\pi/2$) to that at the pole ($\theta=0$) is 1+b/a. Guided by the suite of numerical simulations in  \citet{radice2018}, we set this ratio to $\sim 50$ by fixing $a=1$ and $b=50$. 
We assume the mass of the dynamical ejecta ($M$) 
varies with radial ejecta velocity ($v_{\rm ej}$) as 
\begin{equation}
    M(v_{\rm ej}>v)\,\propto v^{-5}.
\end{equation}
This distribution extends from a minimum 
velocity $v_{\rm min}= 0.2c$ up to $v_{\rm max}= 0.8c$. 
All properties of the dynamical ejecta are motivated 
by numerical simulations of BNS mergers \citep{sekiguchi2016,kawaguchi2018,radice2018,hotokezaka2018}. 

\begin{figure}[ht!]
	\plotone{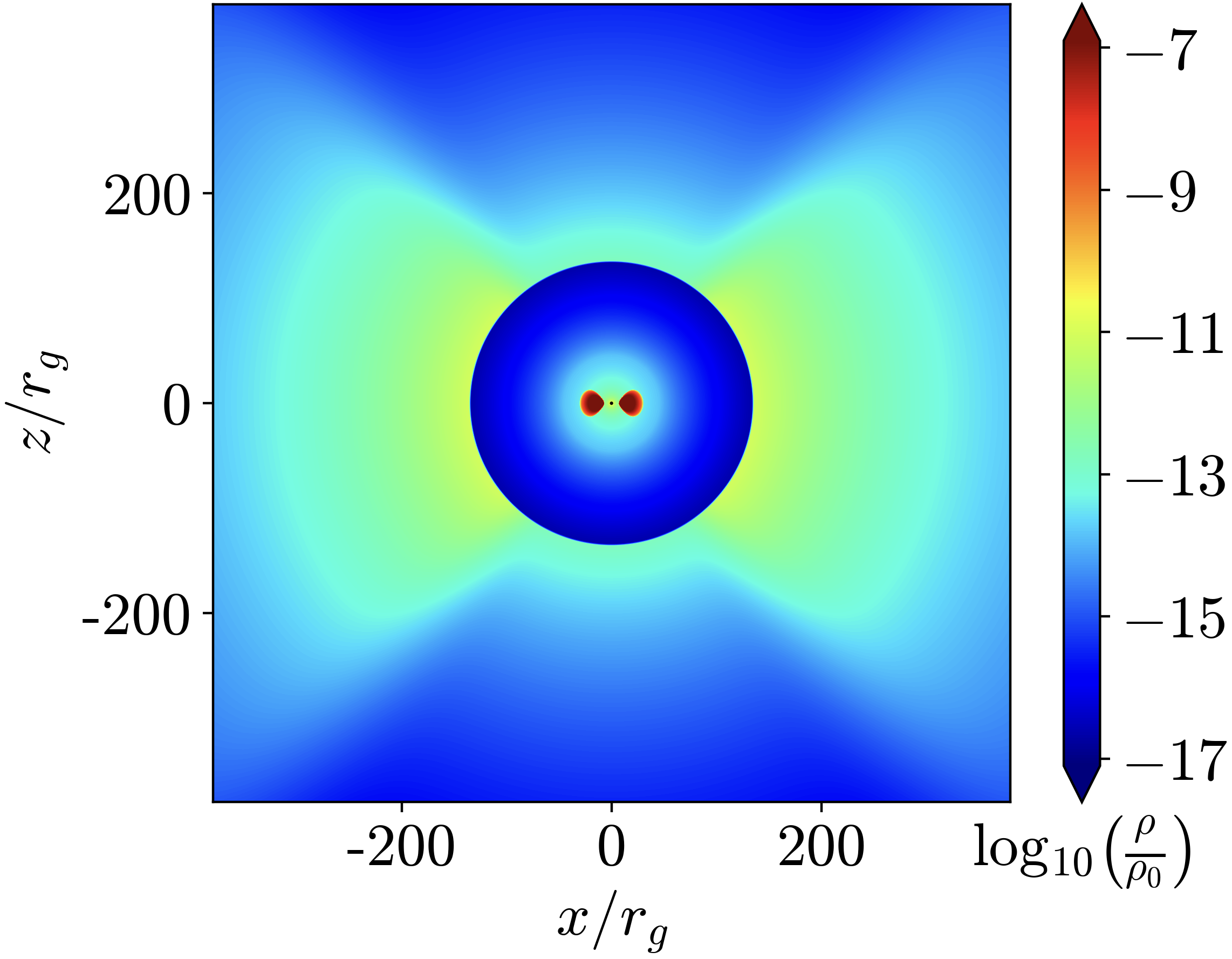}\\
    \plotone{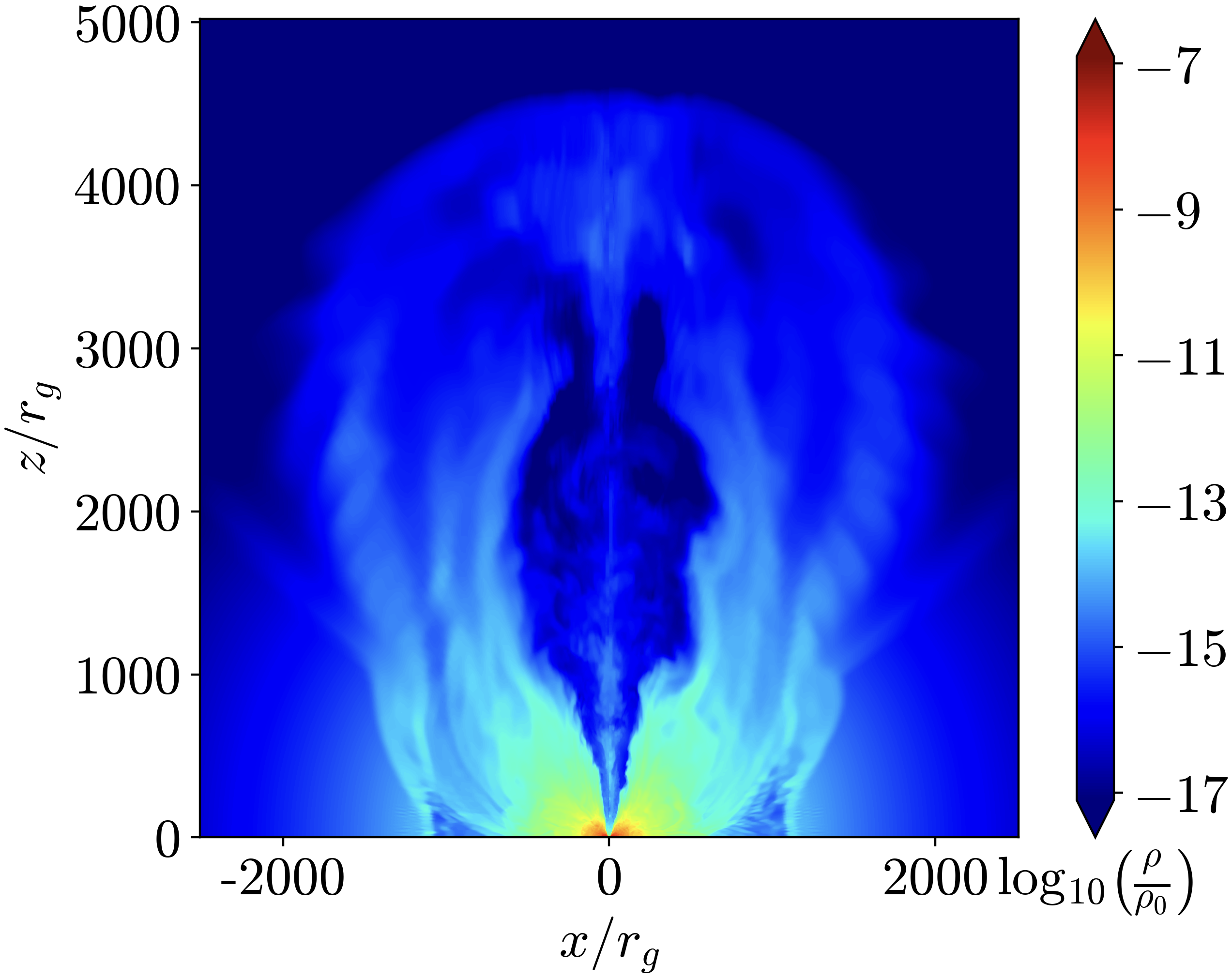}
    \caption{Density contour plots showing the initial setup (top panel) of our 3D GRMHD setup, consisting of a black hole surrounded by a magnetized accretion disk and dynamical ejecta, where $\rho_0 = 7\times 10^{16}$ g cm$^{-3}$. Bottom panel: snapshot at $\sim 0.08$ s, accretion onto the black hole launches jets and disk winds along the z-direction.}
    \label{rho_snapshot}
\end{figure}

The top panel of Fig. \ref{rho_snapshot} shows the density contour plot of our initial setup, with the central region consisting of a black hole-torus system surrounded by the dynamical ejecta. The bottom panel shows a snapshot at 0.08 seconds, after the onset of accretion, where a jet surrounded by disk winds is launched from the central engine.

\subsection{Evolving the blast wave beyond the disk}
\begin{figure}
	\includegraphics[width=\columnwidth]{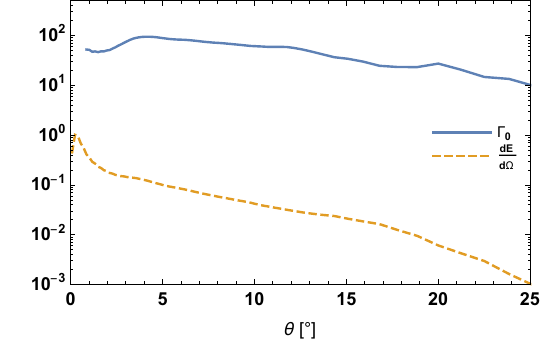}
 
    \caption{Azimuthal and time averaged Lorentz factor (solid blue line) and energy per solid angle distribution of the jet (in units of $G=M=c=1$, and solid angle in steradians), with peak normalized to 1 (dashed orange line) from GRMHD simulations, extracted at $\sim 2000$ r$_{\rm g}$. This profile is used as the initial conditions to evolve the jet through an AGN disk like external medium.}
    \label{jetstructure}
\end {figure}

We extract the time and azimuthally averaged properties  (Lorentz factor and energy) of the 
outflow from the 3D GRMHD simulations at $\sim 2000 r_{g}$, starting from $10^{-5}$ s to 1.5 s. These averaged profiles are used as initial conditions in our next step. 
The Lorentz factor and normalized energy distribution of the 
jet (averaged over time) are shown in Fig. \ref{jetstructure}, where the quantities 
are plotted against the polar angle $\theta$. To evolve the 
blast wave to larger radii, we use equation (5) in \cite{peer2012},
\beq
\frac{d\Gamma}{dm} = -\frac{\hat{\gamma}(\Gamma^2-1)-(\hat{\gamma}-1)\Gamma\beta^2}{M+m(2\hat{\gamma}\Gamma-(\hat{\gamma}-1)(1+\Gamma^{-2}))} \,,
\label{LFevolution}
\eeq
which governs the Lorentz factor evolution of a blast wave 
(assuming it is adiabatic), propagating through an arbitrary 
density distribution, taking into account the momentum 
and thermal energy contribution of the shocked material 
to the energy of the blast wave. Here, $\Gamma$ is the 
bulk Lorentz factor of the blast wave, 
$\beta$ is the velocity of the blast wave as a fraction of 
light speed, $m$ is the mass swept up by the blast wave, 
$\hat{\gamma}$ is the adiabatic index (we use 4/3 as the AGN
disk is likely to be radiation pressure dominated where
we envision BNS merger events), 
and $M$ is the mass of the outflow, which, from our 
simulations, has a rest mass energy of $\approx 5\times 10^{50}$ erg.

We propagate the blast wave through a Gaussian density 
distribution, as expected in AGN disks (see \S \ref{abs} 
for more details on the density used), assuming the merger occurs at the disk mid plane, and the blast wave travels perpendicular to the mid plane. $m$ is calculated 
by integrating the external density distribution over a projected 
area subtended by the solid angle of the blast wave,
\beq
m = \int_{0}^{2\pi}\int_{0}^{\theta_j}\int_{r_0}^{r_{\rm out}}{\rho_{\rm AGN}\,r^2}\,{\rm sin}\theta\,dr\,d\theta \,d\phi \, ,
\label{sweptmass}
\eeq
where $\rho_{\rm AGN}$ is the mass density of the AGN disk (see Sec. \ref{abs} for details about the density distribution), $r$ is the radial distance
of the blast wave from the merger source, starting from a radius of $r_0 = 2000$ r$_{g}$ out to $r_{\rm ext}$, which is fixed by setting the maximum observing time to 1000 days. $\theta_j$ is the angular extent of the blast wave ($\sim 15^{\circ}$), while the relativistic jets do not contain much of the dynamical ejecta, the ejecta plays a part in collimating and shaping the distribution of the outflow (see Fig. \ref{rho_snapshot}, bottom panel).
To implement equation \ref{LFevolution}, the blast wave is divided 
into a 50 $\times$ 50 grid in $\theta$ and $\phi$, 
and each patch is evolved in time by integrating 
equation \ref{LFevolution} using the 4th order Runge-Kutta 
method, to get the Lorentz factor as a function of time. 
The solid angle of each patch is $\sim 10^{-4}$, 
which is equal to or smaller than $\frac{1}{\Gamma_0^2}$, 
where $\Gamma_0$ is the initial Lorentz factor 
of the jet and has a peak value of $\sim 100$ 
(see Fig. \ref{jetstructure}). This ensures the patches 
are small enough so that beaming effects from each 
patch will be adequately resolved when calculating 
the observed emission.

\section{Calculating the Afterglow emission}

With the dynamics (energy and velocity vs time) of the 
blast wave known, we can now obtain the non-thermal 
afterglow spectrum of the shocked material, as the outflow 
propagates through the disk. The foundational paper 
which calculated synchrotron emission from afterglows 
is \citep{sari1998}, however, this assumes a highly relativistic 
blast wave. Later works such as \citep{granot2002,leventis2013} 
provide the prescription to calculate the afterglow emission 
for blast waves ranging from highly relativistic to non-relativistic 
regimes, and they also include additional modifications in the synchrotron 
spectrum due to synchrotron self-absorption 
and fast cooling. We use these prescriptions, with the 
dynamics provided by our model to obtain light curves and spectra. 
In general, the non-thermal electron population ($N$) is 
assumed to be a power-law distribution as function of 
the electron Lorentz factor ($\gamma_{\rm e}$), i.e., 
$dN/d\gamma_{\rm_e}\propto\gamma_{\rm e}^{-p}$,  with a minimum Lorentz factor
\beq
\gamma_{min}= \frac{\epsilon_e m_p}{\xi m_e}\frac{p-2}{p-1}\Gamma,
\eeq
where $\xi$ is the fraction of electrons accelerated into the non-thermal spectrum, assumed to be 15\% \citep{guo2014}, $m_p$, $m_e$ are the proton and electron masses respectively. $\epsilon_e$ is the fraction of the blast wave energy supplied to the non-thermal electrons. The maximum Lorentz factor of the distribution depends on the cooling rate of the synchrotron electrons (\citealt{eerten2010}).
The resulting afterglow spectrum takes the form 
$F_\nu\propto \nu^\alpha$, where $F_\nu$ is the 
spectral flux and $\alpha$ is either a constant, or 
a function of the non-thermal electron  power law slope ($p$). 
Since we divide the surface of the blast wave into many 
patches, each with its own velocity and energy, 
our model does not assume spherical symmetry for 
the whole blast wave when calculating the resulting 
afterglow emission. 

A few assumptions have to made about the blast wave's parameters when calculating the synchrotron emission. The micro-physical parameters $\epsilon_B$ and $\epsilon_{e}$ are the fraction of the blast wave's energy that is converted to magnetic energy density and energy of the accelerated electrons in the forward shock, respectively. These two parameters, along with the spectral index ($p$) and the GRB prompt efficiency are required for the calculation. It is widely believed that collisionless shocks are responsible for GRB afterglows \citep{Katz2007}. For observed GRBs, which occur in interstellar media, this is a valid model. We will show that the collisionless shock model also holds for afterglows in regions of the AGN disks, that are relevant to our calculations. Thereby enabling us to use the parameters inferred from observed GRB afterglows. 

The mean free path for Coulomb collisions ($\Lambda_{\rm mfp}$) in a forward shock is $\Lambda_{\rm mfp}\sim 1/\Gamma n \pi d^2 \sim 10^{31}\Gamma (1\,{\rm cm}^{-3}/n)$ cm \citep{waxman2006}, where $\pi d^2$ is the cross-section for Coulomb collisions, with $d\sim e^2/\Gamma m_p c^2\sim 10^{-16}\Gamma^{-2}$ cm. \,$\Gamma$ is the Lorentz factor of the blast wave, $e$ is the charge of an electron, $m_p$ is the proton mass, c is the speed of light, and $n$ is the external number density. In our calculations, we consider the afterglow emission originating from regions where the scattering optical depth ($\tau_{\rm scatt}$) is less than 1 (see Sec. \ref{abs} for more details). 
We can approximate $\tau_{\rm scatt}\sim n\sigma_T H$, where $\sigma_T$ is the Thomson scattering cross section and $H$ is the AGN disk scale height, which we use as an estimate for the physical size of the system under consideration. The condition $\tau_{\rm scatt}<1$ yields $n\lesssim 10^{24} H^{-1}$ cm$^{-3}$. 
With this, we obtain $\Lambda_{\rm mfp}\gtrsim10^7\Gamma (H/ 1\,\rm{cm})$ cm, substituting a modest Lorentz factor of 10 yields a mean free path that is $10^8$ times greater than the physical size of the system. Therefore, the collisionless shock model is viable in this system as well.
$\epsilon_B$ is proportional to $B^2/n$ (assuming the initial Lorentz factor is similar across short GRBs), where $B$ and $n$ are the magnetic field strength and number density of the external medium respectively. From modeling observed short GRB parameters, this ratio has a range $\sim 10^{-6} - 10^{-1}$ (\citealt{fong2017}). AGN disks have inferred magnetic field strengths of $\sim 10^4$ G (\citealt{daly2019}), and AGN disk models (\citealt{sirko2003,thompson2005}) predict densities ranging from $\sim 10^{8} -10^{15}$ cm$^{-3}$. Hence, it is reasonable to assume GRBs in AGN disks have $\epsilon_B$ values similar to those inferred from observed GRBs, which occur in interstellar media. 

With the above considerations, we use typical values of $\epsilon_B = 10^{-4}$, $\epsilon_e = 10^{-2}$ and $p = 2.17$, obtained from modeling observed GRBs \citep{margutti2018,troja2018}. 
The total energy of each jet produced in the simulation 
is $\sim 10^{51}$ erg, which is typical of observed GRBs. 
Some of this energy will power the prompt emission, this efficiency (the 'prompt efficiency') varies between a few
percent to more than 90\% \citep{fong2015}. Since the GRB prompt emission is attributed to an internal mechanism within the jet, at very early stages of the jet evolution, within distances of $\lesssim 10^{11}$ cm (e.g., \citealt{beloborodov2017,giannios2008}), we can expect a similar prompt efficiency from jets within AGN disks. Therefore, we will assume a middle value of $50\%$ of the jet's energy powers the 
afterglow. However, see \citealt{lazzati2022}, who find that in the case where the prompt emission is powered by synchrotron emission in reverse shocks, the prompt efficiency in AGN disks can be greater than 90\%. Since the afterglow's peak luminosity is proportional to the energy of the blast wave, such high prompt efficiencies would reduce the afterglow luminosity by a factor $\sim 2$. We also assume
that each patch independently propagates radially away from the source.
Next we discuss how absorption and 
scattering of photons is handled as they 
propagate through the AGN disk.

\subsection{Absorption and Scattering}
\label{abs}

As a beam of photons propagate through a medium, 
it can be absorbed and scattered by the medium, leading to 
an attenuated beam and diminished emission. 
Consequently, some of the absorbed emission will be 
reprocessed by the medium, this emission will depend 
on the properties of the medium and the absorption 
mechanism. Photons that get scattered, e.g., 
elastic scattering in the simplest case, will just change 
their direction of propagation, which affects the amount 
of photons and time taken to reach an observer. 
The equation of radiation transfer, considering only absorption, can be written as (\citealt{rybicki}),
\beq
\frac{dI_{\nu}}{ds} = -\alpha_{\nu}I_{\nu} + j_{\nu} \, ,
\label{radtrans}
\eeq
where $s$ is the path from the emission region to the 
observer, $I_{\nu}$ is the frequency dependant intensity (derived from the non-thermal synchrotron flux $F_{\nu}$) 
from the emitter, $\alpha_{\nu}$ is the absorption coefficient 
and $j_{\nu}$ is the reprocessed emission from the 
background medium. The first term on the right hand side 
describes the attenuated emission from the emitter 
(the blast wave in our case) along the line of sight, 
and the second term is the reprocessed emission, 
attenuated along the line of sight. For the disk temperature considered in our observability calculations ($10^4$ K), the AGN disk opacity is dominated by free-free absorption by hydrogen and electron scattering \citep{sirko2003,thompson2005}, hence we will only account for these two mechanisms in our absorption calculations. The absorption coefficient for free-free absorption takes the form \citep{rybicki},
\beq
\alpha_{\nu, \rm{ff}} = 3.7\times10^8T^{-\frac{1}{2}}Z^2n_en_i\nu^{-3}(1-e^{-\frac{h\nu}{kT}})g_{\rm{f}}\,\,\,\, {\rm cm^{-1}}\, ,
\label{ffabs}
\eeq
where $T$ is the temperature, $Z$ is the atomic number, $n_e$ is the electron number density, $n_i$ is the ion number density, all associated with the background medium (the absorber). $g_{\rm{f}}$ is the Gaunt factor (which will be fixed to 1.2 in our applications), $h$ is the Planck's constant and $k$ is the Boltzmann constant. We assume $n_e = n_i$ and fix $Z=1$, taking into account only the hydrogen component, which dominates the disk composition.

In all our observability constraints, we compare the emission from the afterglow to the observed quiescent AGN emission. Since the quiescent emission is generated by the entire disk, and the reprocessed emission in the radiation transfer calculation (second term) primarily originates from portions of the disk from the blast wave along the line of sight, it will be negligible in comparison to the quiescent disk emission. In other words, for equation \ref{radtrans}, the integral of $j_{\nu}$ along the line of sight will be overwhelmingly dominated by the observed quiescent disk emission.

\begin{figure}
	\includegraphics[width=\columnwidth]{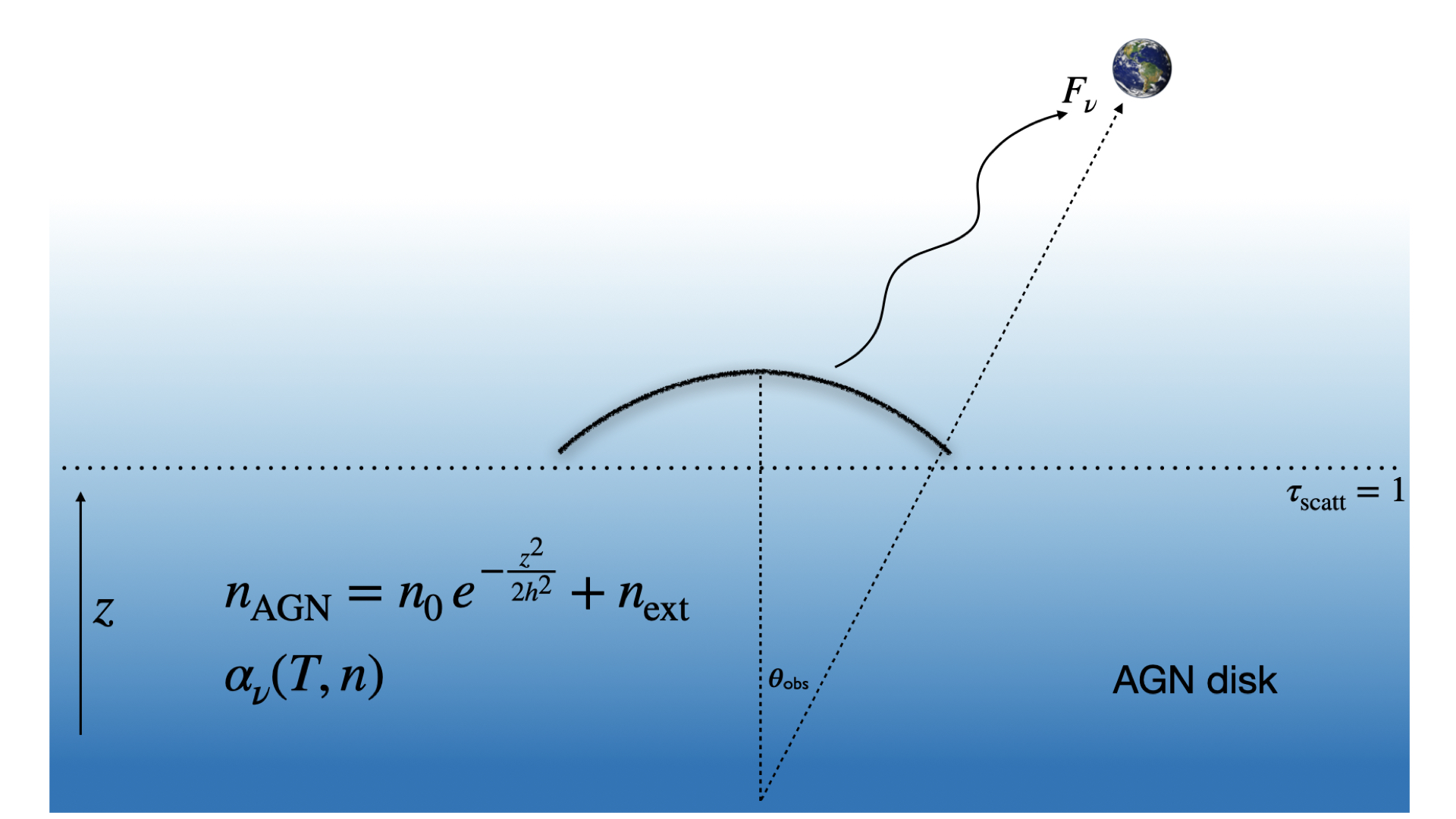}
    \caption{A sketch of the setup envisioned. While the blast wave is below the $\tau_{\rm scatt} = 1$ surface within the AGN disk, we assume no emission from the blast wave reaches the observer. Beyond the $\tau_{\rm scatt} = 1$ surface, the synchrotron emission ($F_{\nu}$) is calculated as it propagates through the AGN disk to the observer, taking into account attenuation due to absorption (with absorption coefficient $\alpha_{\nu}$, see \S \ref{abs} for details).  The emission is summed across the entire blast wave, taking into account photon travel time and relativistic effects.}
    \label{schematic}
\end{figure}

The background density profile ($n$) of an AGN disk is 
described in \citep{sirko2003,thompson2005,Dittmann2020}. 
In all these works, the vertical density distribution is Gaussian, although depending on the radial distance from the SMBH, the disk may be supported by gas or radiation pressure, leading to different density profiles.
Since we evolve the blast wave to beyond the disk scale height, 
we have to include the density external to the disk as well 
($n_{\rm ext}$). We will assume the external density is 
constant and add the two density distributions to ensure 
a smooth transition from the disk to the external medium. 
Hence the density is of the form
\beq
n_{\rm AGN} = n_0 e^{-\frac{z^2}{2H^2}} + n_{\rm ext}
\label{densityprofile}
\eeq
where $n_0$ is the central density at the 
mid-plane (which depends on the distance to the 
central SMBH). $H$ is the scale height of the disk and 
$z$ is the direction perpendicular to the mid-plane. 
$H$ and $n_0$ can be obtained from the various 
AGN disk models for a fixed set of parameters pertaining 
to the AGN, e.g., SMBH mass, accretion rate, and 
distance from the SMBH. In our light curve and spectrum examples (Figs. \ref{tempSED}, \ref{1e14LC}, \ref{angularprofile}), we fix $H = 3\times10^{13}$ cm and $n_0 = 3\times10^{12}$ cm$^{-3}$, while surveying over $H = 10^{13}-10^{18}$ cm and $n_0 = 10^{6}-10^{15}$ cm$^{-3}$ when investigating the detectability of the afterglow (Fig. \ref{observability}). For each calculation, we ignore the radial variations of $n_0$ as the blast wave propagates out, since the radial extent of the blast wave is much smaller than the distance traversed.  In addition, the temperature 
$T$ of the disk, as a function of distance from the 
central SMBH is also given in these models 
(which will be needed to calculate absorption effects). 
In all our calculations, we assume the progenitor is 
located at the disk mid-plane, and the jet propagates 
perpendicular to this plane, along the $z$ direction (see Fig. \ref{schematic} for a pictorial description of this setup).  This configuration reduces the number of free parameters in our calculations, allowing us to clearly study how the observability depends on the disk parameters (scale height and density). Additionally, studies involving numerical simulations of binary BHs in AGN disks find their orbit tend to settle along the mid-plane of the disk (\citealt{Dittmann2023}).
The value of $n_{\rm ext}$ is fixed at 100 cm$^{-3}$,
given that $n_0$ often exceeds $10^{12}$ cm$^{-3}$,
our results are insensitive to the choice
of $n_{\rm ext}$ as long as it is small in comparison. The intensity in equation \ref{radtrans} is integrated from the position of the blast wave (when $\tau_{\rm scatt}<1$), out to a distance of $z=10H$

Treating scattering is much more difficult,  photons change their paths as they propagate, they can be absorbed and change their wavelength as they undergo multiple scatterings, so a radiation transport code is required to accurately asses the effects of scattering (see \citealt{wang2022} for a study on the effects of scattering on embedded afterglows). Radiation transport is computationally intensive and hinders a parameter space study of AGN-embedded systems. Here we implement a crude calculation to take into account the effect of scattering as follows. At each time step, we calculate the scattering optical depth integrated along the line of sight from the blast wave to the observer, which is given by,
\beq
\tau_{\rm scatt} = \int_{r_{0}}^{r_{\rm edge}} n \sigma_T ds
\eeq
where the integral is carried out from the location of the blast wave element ($r_{0}$) to the edge of the disk ($r_{\rm edge} = 10H$) along the line of sight ($s$). $n$ is the number density along $s$ and $\sigma_T$ is the Thomson scattering cross section. 
If $\tau_{\rm scatt}>1$, we set the emission from this portion of 
the blast wave to 0. 
If $\tau_{\rm scatt}\le1$, the photons will most likely not 
undergo a scattering before they reach the observer, 
so in this case we assume all the emission reaches the observer \footnote{Changing the scattering optical depth condition to a more restrictive one of $\tau_{\rm scatt}\le0.3$ leaves our conclusions largely unchanged, due to the fact that $\tau_{\rm scatt}$ is either $\ll1$ or $\gg1$ for a majority of the time as the jet propagates through the disk.}. 
The energy of the scattering electrons within the AGN disk is equivalent to $1.2\times 10^{20}$ Hz, which is much higher than the highest frequency ($10^{18}$ Hz) considered in our light curves and spectra. Therefore, Thomson scattering is a good approximation in this situation. For large optical depths, the scattered emission will emerge on a diffusive timescale of $t_{\rm diff}\simeq 2550\,(H/10^{14})^{2}(n_{0}/10^{13})$ days, with this diffuse emission attenuated by a factor ($t_{\rm diff}/T_{\rm lc})(\Omega_j/4\pi)$ (\citealt{wang2022}), where $\Omega_j\approx0.02$ is the solid angle subtended by the jet, and $T_{\rm lc}$ is the characteristic timescale of the unobscured light curve. 
The unobscured light curves typically peak less than a day for on-axis observers (see Figs. \ref{tempSED}, \ref{1e14LC}), which means the diffuse emission due to scattering will be suppressed by $\sim 3-5$ orders of magnitude, and will appear at much later times compared to the emission from the regions where $\tau_{\rm scatt} <1$, consistent with the findings of \citealt{wang2022}. These two properties will render the scattered emission difficult to observe above the quiescent emission of the AGN, and will make it difficult to associate an EM counterpart with a coincident GW detection.

Using equations \ref{densityprofile} and \ref{sweptmass}, 
we can obtain the swept up mass ($m$) in terms of the 
radius of the blast wave ($R$), which allows us to express 
the Lorentz factor evolution (equation \ref{LFevolution}) 
in terms of $R$. We can then solve this differential equation to 
get the velocity and energy of the blast wave in terms of radius. 
These dynamic quantities are used to calculate the 
afterglow flux in the co-moving frame of the shocked fluid. 
We assume the emitted flux in the co-moving frame is 
isotropic and boost it to the observer frame. 
The observed flux as a function of observed time 
($T_{\rm obs}$) is obtained via the relation,
\beq
T_{\rm obs}=\int{\frac{dR}{\beta c}(1-\beta\,\cos\alpha_s)}\, , 
\eeq
where $\alpha_s$ is the angle between the velocity vector 
of a blast wave segment and its line of sight towards the observer, 
$\beta$ is the ratio of the velocity of the blast wave to the 
speed of light, and $c$ is the speed of light. 
The emitted flux can then be plugged into equation 
\ref{radtrans} and integrated along the line of sight to 
obtain the observed attenuated emission.

\section{Spectra, Light Curves and observability}

We present the observable quantities, synthetic light 
curves and time dependent spectra, for the afterglow. 
The prospects of detecting this transient, when considering 
the sensitivity of the detector and quiescent emission of 
the AGN is discussed as well. We explore the detectability of 
these embedded systems, for a range of AGN disk heights 
and central densities.

\subsection{Spectral evolution}

In Fig. \ref{tempSED}, we present the time evolution of 
the afterglow spectra for a jet embedded within a disk of 
scale height $3\times 10^{13}$ cm and central density 
$3 \times 10^{12}$ cm$^{-3}$, which corresponds to a system located $\sim 5000$ r$_g$ away from a $10^8\,M_{\odot}$ SMBH under the \citealt{thompson2005} AGN disk model. This parameter set admits an observable case for the afterglow (see Sec. \ref{survey} and Fig. \ref{observability}). We assume the post-merger system is located at the mid-plane of the AGN disk. 
The quiescent AGN emission of M87 (from \citealt{M87spectrumEHT}) and NGC 4151\footnote{Data obtained from \href{https://ned.ipac.caltech.edu/}{
The NASA/IPAC Extragalactic Database}, see references therein.} (assumed to be steady in time) is also 
included as well-observed examples of the quiescent emission from a low-luminosity AGN and Seyfert galaxy respectively. All emitting sources are scaled to a distance of 50 Mpc (as a representative distance to possible LIGO sources). 
The top panel is for an observer on-axis 
($\theta_{\rm obs} = 0^{\circ}$) with respect to the jet and 
the bottom panel is for an observer $30^{\circ}$ off-axis. 
We see that in the on-axis case, the afterglow outshines 
the quiescent emission above $\sim 10^{12}$ Hz for about 
a day and then fades. For the off-axis case, the afterglow 
does not outshine the quiescent emission and will therefore 
not be detectable. For the on-axis case, the spectrum is 
initially self-absorbed at low frequencies with 
$F_{\nu}\propto\nu^{\frac{1}{3}}$. The spectrum rises up 
to the injection frequency of the electrons (the synchrotron frequency of electrons at $\gamma_{min}$) and then 
decays as a power law proportional to $(1-p)/2\sim -0.6$. 
As the blast wave slows down, the injection frequency 
decreases, causing the peak to shift to lower frequencies 
while the flux gradually decreases. For the off-axis case, 
the rise in the spectrum is delayed (due to beaming effects), 
at these later times, the blast wave has expanded 
considerably to the point where self-absorption does not 
play an important role in the spectrum. For these times, 
the off-axis observer only sees the power-law spectra 
rise and fall as the blast wave decelerates. 

This figure elucidates another important outcome regarding 
observability. At lower frequencies, absorption effects are 
more pronounced and the disk emission is higher; 
these two effects make it unfeasible to detect afterglows 
roughly below the I-R band. However,  absorption 
effects and disk emission are lower at higher frequencies, 
which enables a greater chance of detecting 
embedded afterglows. 

\subsection{Light curves and angular profile of the afterglow}

\begin{figure}
	\includegraphics[width=\columnwidth]{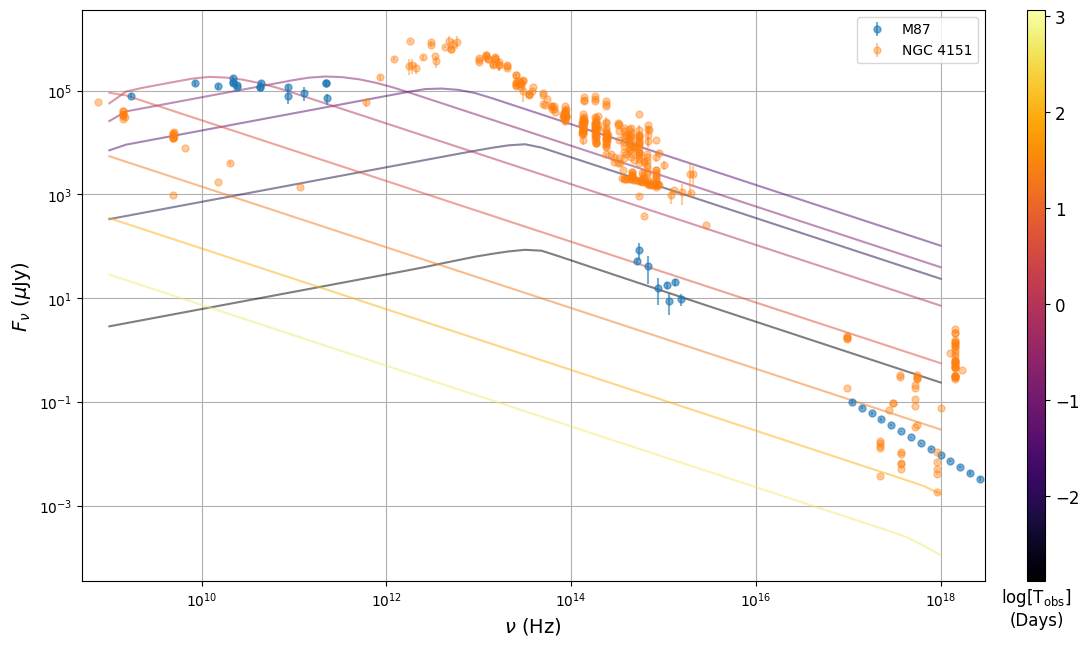}\\
    \includegraphics[width=\columnwidth]{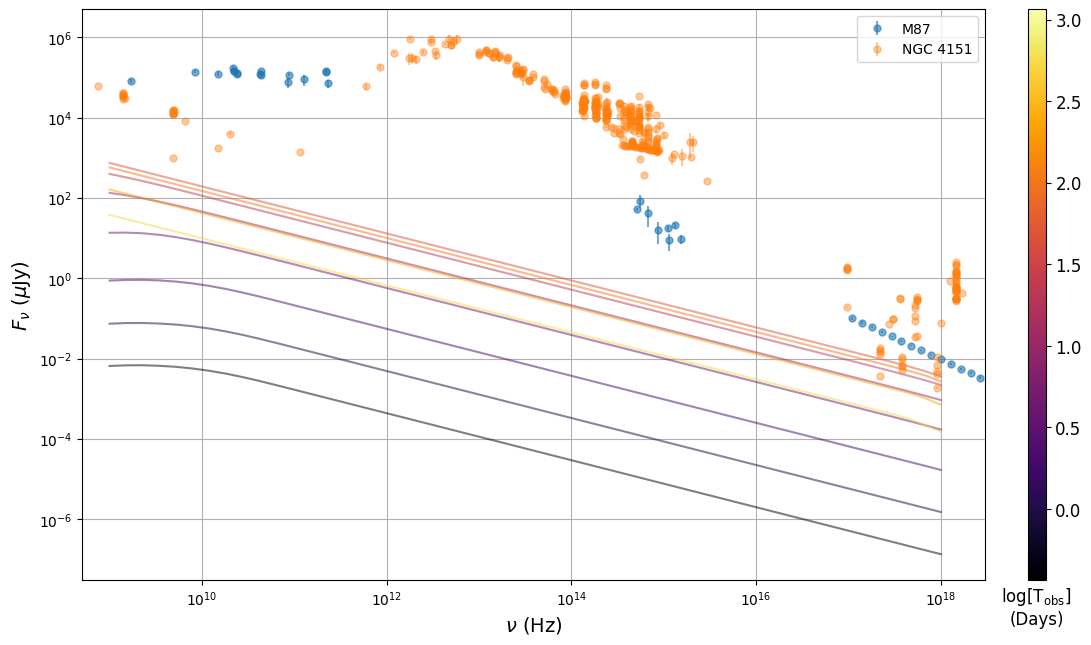}
 
    \caption{Temporal evolution of an afterglow spectrum embedded in a disk scale height of $3\times$ 10$^{13}$ cm, $n_0$ = 3$\times$ 10$^{12}$ cm$^{-3}$ and temperature of 10$^4$ K, corresponding to a merger occuring $\sim 5000$ r$_{\rm g}$ away from a $10^8$ M$_{\odot}$ BH. Top panel is for $\theta_{\rm obs}$ = 0$^{\circ}$ and bottom panel shows the spectrum for $\theta_{\rm obs}$ = 30$^{\circ}$, the color bar shows observer time in days. The spectra of M87 \citep{M87spectrumEHT} and NGC 4151 are included. For the on-axis case, the afterglow outshines M87 from $\sim 0.1-1$ days at frequencies greater than $\sim 10^{14}$ Hz. The afterglow does not outshine M87 for the off-axis case and hence will not be detectable. Due to absorption effects and higher quiescent emission at lower frequencies, the afterglow has a better chance of being detected at higher frequencies.}
    \label{tempSED}
\end{figure}

\begin{figure}
    \includegraphics[width=\columnwidth]{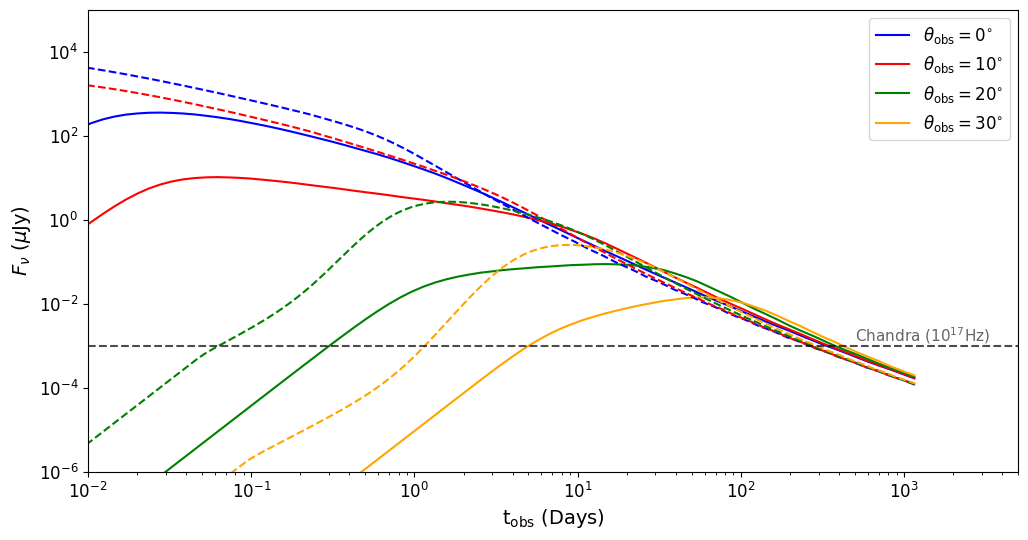}\\
	\includegraphics[width=\columnwidth]{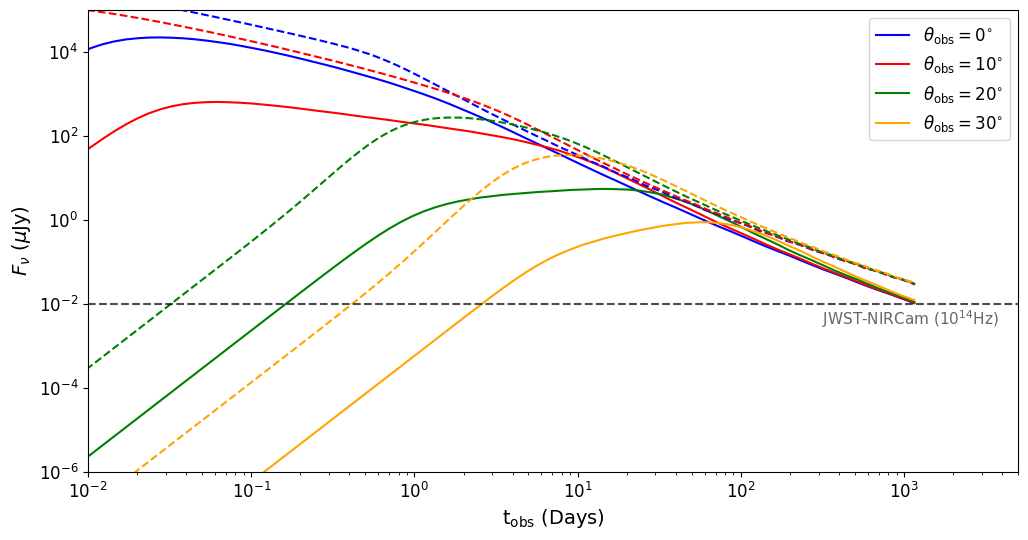}
    \caption{Synthetic light curves at $10^{17}$ Hz (Top panel) and $10^{14}$ Hz (Bottom panel) for various observing angles, along with the sensitivity limit of of Chandra ($\sim 10^4$ s exposure time) and JWST-NIRCam ($\sim 10^3$ s exposure time). The solid lines are obtained from the structured jet from GRMHD simulations. For comparison, the afterglow from a top-hat jet is showed with dashed lines with an opening angle of $10^{\circ}$, initial Lorentz factor of 100 and total energy of $5\times 10^{50}$ ergs. The top-hat jet afterglow is brighter due to the larger Lorentz factors and energy per solid angle of the jet at late times. Although the sensitivity limits indicate that the afterglow is detectable for observers beyond $30^{\circ}$, the emission from M87 lies at $\sim 10^{-1}$ $\mu$Jy and $\sim 10^{2}$ $\mu$Jy at $10^{17}$ Hz and $10^{14}$ Hz respectively. This restricts the afterglow observability to near on-axis observers.}
    \label{1e14LC}
\end{figure}

Fig. \ref{1e14LC} shows the light curves for the same 
conditions as in Fig. \ref{tempSED}, for various 
observing angles at a frequency of $10^{17}$ Hz (top panel) and 
$10^{14}$ Hz (bottom panel). The sensitivity limit of Chandra and JWST 
for these observing frequencies are also included. 
The solid lines show the synthetic light curve of the 
structured jet obtained from our GRMHD simulation. 
For comparison, we also show the light curves from 
a top-hat jet (dashed lines), with uniform properties within the opening angle, for the various observing 
angles. The jet has an opening angle of $10^{\circ}$, 
initial Lorentz factor of 100 (which is the maximum 
Lorentz factor of the structured jet), and a total 
energy of $5\times10^{50}$ ergs distributed uniformly across the jet. The flux from 
this top-hat jet is higher than the structured jet due to the 
higher energy per solid angle and larger Lorentz factor 
towards the edge of the jet. Therefore, a top-hat jet model 
may overestimate the afterglow observability of these 
systems.  Even though the instrument sensitivity is 
capable of detecting afterglows for observers $30^{\circ}$ 
off-axis, the quiescent emission of M87 restricts the 
detectability to near on-axis observers 
(within $\sim 20^{\circ}$) (see Fig. \ref{tempSED}). 

\begin{figure}
    \includegraphics[width=\columnwidth]{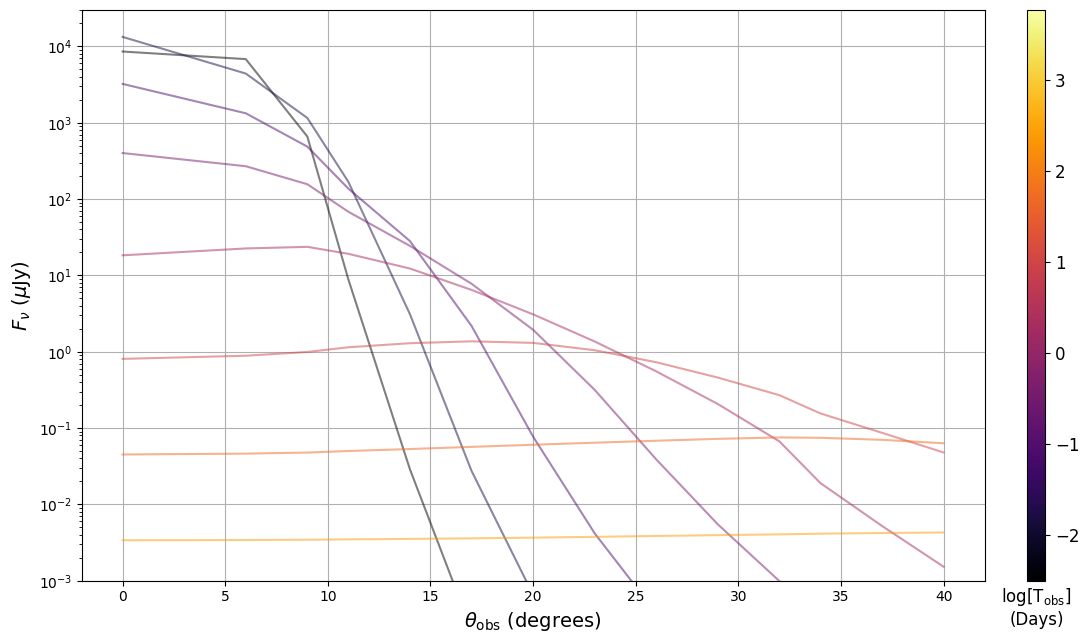}\\
	\includegraphics[width=\columnwidth]{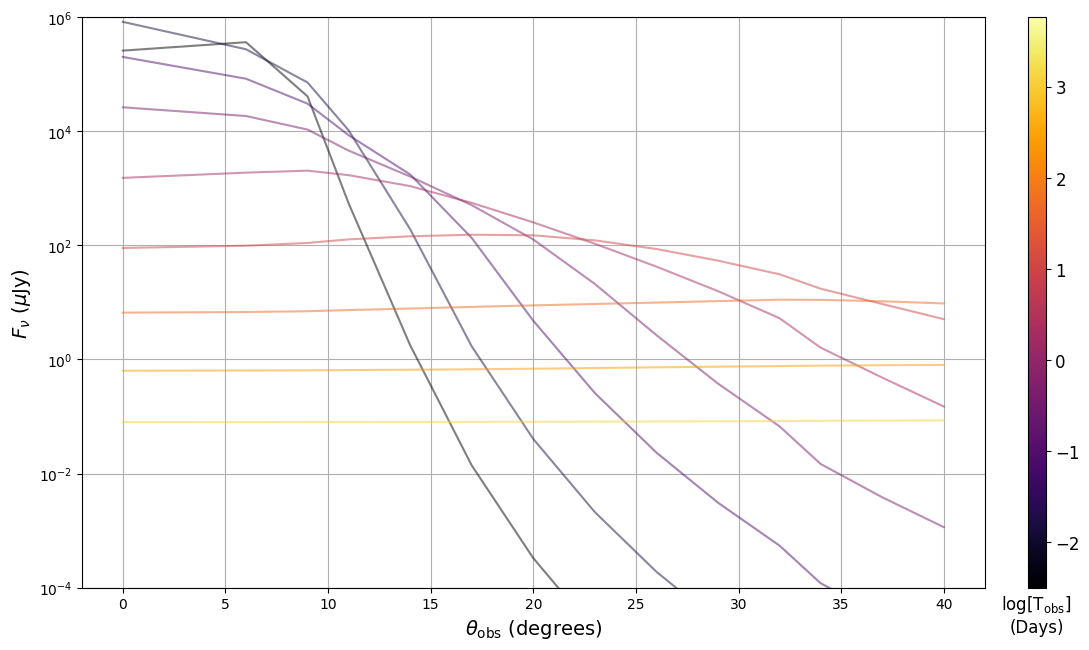}
    \caption{Temporal evolution of the afterglow flux vs. observing angle at $10^{17}$ Hz (top panel) and $10^{14}$ Hz (bottom panel), for the same initial and external conditions as in Figs. \ref{tempSED} and \ref{1e14LC}. At early times, the emission is concentrated close to the jet axis due to beaming effects and higher energy per solid angle. As the blast wave decelerates, the flux at larger angles increases over time due to diminished beaming effects. Once the blast wave becomes non-relativistic, the emission is spherically symmetric, which is portrayed by the nearly horizontal lines at late times.}
    \label{angularprofile}
\end{figure}

For completeness, we also show the time evolution 
of flux vs. observing angle in Fig. \ref{angularprofile}. 
At early times, the emission is highly concentrated 
along the jet axis ($\theta_{\rm obs} = 0^{\circ}$) 
due to strong beaming resulting from the higher 
Lorentz factors near the center. The edges 
($\gtrsim 10^{\circ}$) have a much lower Lorentz 
factor and energy per solid angle (see Fig. \ref{jetstructure}, 
for the initial Lorentz factor and energy per solid 
angle distribution). As the blast wave decelerates, 
beaming diminishes and  the flux at larger observing 
angles starts to increase. Once the blast wave becomes 
non-relativistic, the emission is essentially spherical 
and the observed flux will be the same for any 
observing angle.

We note that for the AGN disk height ($3\times 10^{13}$ cm) 
and central density ($3 \times 10^{12}$ cm$^{-3}$) used here, 
the $\tau_{\rm scatt} =1 $ surface is at $\sim 5\times10^{11}$ cm. 
So the emission is generated starting from the inner regions of the disk, 
the blast wave is still relativistic, 
with a peak Lorentz factor of $\sim 70$ at this distance.


\subsection{ A parameter space survey for afterglow observability}
\label{survey}

\begin{figure}
	\includegraphics[width=\columnwidth]{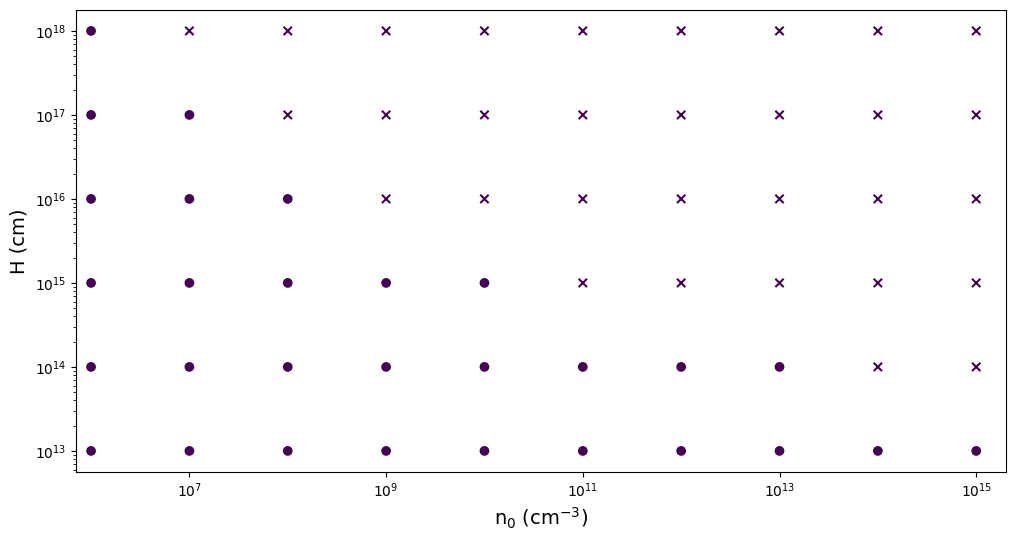}
    \caption{Detectability of afterglows (for on-axis observers) for varying disk heights (H) and central densities ($n_0$). If the afterglow is brighter than M87 SED, it is detectable (dots), if it is always dimmer than M87 SED, it is not detectable (crosses). The disk temperature is fixed at $10^4$ K. }
    \label{observability}
\end{figure}

Next, we perform a parameter space survey for the AGN disk 
properties and assess the detectability of the afterglow. 
We conduct our parameter survey 
over the physical properties, disk height and central densities, 
so that our results can be applied to a set of disk models of one's choosing.  
Fig. \ref{observability} shows the observability of afterglows, 
above the quiescent emission of M87, for various disk heights 
($H$) and central densities ($n_0$), 
keeping the temperature fixed at $10^4$ K. 
If the afterglow is brighter than the SED of M87 at 
any point in time, and at any wavelength, 
we categorize it as observable and denote it as a circle 
in the plot. If the afterglow is always dimmer than M87, 
we categorize it as unobservable and denote it as 
an `$\times$' for these sets of parameters. 
These calculations are done for an on-axis observer. 
As expected, for denser and larger disks (upper-right region of Fig. \ref{observability}), the 
emission will not be detectable due to a combination 
of absorption effects and faster deceleration of the blast wave.

From Fig. \ref{observability}, we estimate that for AGN disk column densities ($H\times n_0$) less than $\sim 10^{24}$ cm$^{-2}$, the afterglow will be detectable for the specified parameters of the blastwave. Whereas for column densities larger than $\sim 10^{28}$ cm$^{-2}$, the afterglow will not be detectable.

\section{Conclusion}

We have investigated the detectability of observing 
BNS merger afterglows embedded in AGN disks. 
Our setup is initiated at the horizon scale of the 
progenitor system, extracts the outflows as they are
generated self-consistently from an  accreting compact object, 
and extends the outflows to distances beyond the AGN disk
scale height. The higher ambient densities within an 
AGN disk causes the blast wave to decelerate much 
faster when compared to an isolated environment. 
In some cases, the blast wave becomes non-relativistic 
before it escapes the disk. As a result, the afterglow 
emission will decline a lot faster and have a lower flux, 
the additional absorption effects can hinder the detectability 
of this non-thermal emission. 
However, we find there exists a parameter space (where the AGN disk surface density $\lesssim 10^{24}$ cm$^{-2}$),
 for which the afterglow from a BNS merger could manage 
to outshine the quiescent
disk emission. In particular, observing at
higher frequencies ($\gtrsim 10^{14}$ Hz) 
is more feasible for detection due to lower absorption 
and lower background AGN emission at these frequencies. 
The quiescent AGN emission also restricts the detectability 
of the afterglow to near on-axis observers. These embedded 
afterglows may manifest as a fast variability in the 
quiescent AGN emission, on a timescale of a day, 
at these higher frequencies. We also find that using a 
top-hat jet, with similar bulk energy and velocity properties 
of the structured jet can over-estimate the emission 
compared to the structured jet, due to the higher energies 
and Lorentz factors of the top-hat jet at larger angles.
If pre-merger accretion activities launch outflows, it can inflate a large-scale, low-density bubble 
within the AGN disk, around the merger site. In this scenario, the blast wave could remain 
relativistic throughout its propagation out to the edge of the disk, and 
produce detectable emission, as shown in the
lower-left region of Fig. \ref{observability}.

\begin{acknowledgments}
We are grateful to the anonymous referee for providing constructive feedback. It is a pleasure to acknowledge Eliot Quataert and Daniel Kasen for helpful discussions on the simulation setup. We thank Shengtai Li for assistance with utilizing the HPC resources, and Adam M. Dempsey for helpful discussions. AK acknowledges support from the Director's postdoctoral fellowship funded by LANL LDRD project No. 20220808PRD4. AK, HL and BRR gratefully acknowledge the support by LANL LDRD program
under project No. 20220087DR. Computational resources for this project was
provided by the Los Alamos National Laboratory Institutional 
Computing Program, which is supported by the U.S. 
Department of Energy National Nuclear Security 
Administration under contract No. 89233218CNA000001. This research also used resources of the National Energy Research Scientific Computing Center (NERSC) and the Texas Advanced Computing Center (TACC). NERSC is supported by the Office of Science of the U.S. Department of Energy under Contract No. DE-AC02-05CH11231. This manuscript has been assigned a document release number LA-UR-23-31757.
\end{acknowledgments}
%

\vspace{5mm}





\bibliography{AGN_afterglow}{}
\bibliographystyle{aasjournal}



\end{document}